\documentclass[aps,prl,reprint,superscriptaddress]{revtex4-2}

\usepackage{graphicx}
\usepackage{dcolumn}
\usepackage{bm}

\usepackage{xcolor}
\usepackage{ulem}
\usepackage{siunitx}
\usepackage[inkscapeformat=png]{svg}
\usepackage{soul}

\begin{document}

\preprint{}

\title{Rolling, sliding and trapping of driven particles in square obstacle lattices}

\author{Galor Geva}
\thanks{These authors contributed equally to this work.}
\affiliation{Department of Theoretical Condensed Matter Physics, Universidad Autónoma de Madrid, 28049, Madrid, Spain}
\affiliation{Condensed Matter Physics Center (IFIMAC), Universidad Autónoma de Madrid, 28049, Madrid, Spain}
\affiliation{Instituto Nicolás Cabrera, Universidad Autónoma de Madrid, 28049, Madrid, Spain}
\author{Arin Escobar Ortiz}
\thanks{These authors contributed equally to this work.}
\affiliation{Department of Theoretical Condensed Matter Physics, Universidad Autónoma de Madrid, 28049, Madrid, Spain}
\affiliation{Condensed Matter Physics Center (IFIMAC), Universidad Autónoma de Madrid, 28049, Madrid, Spain}
\affiliation{Instituto Nicolás Cabrera, Universidad Autónoma de Madrid, 28049, Madrid, Spain}
\author{Paula Magrinya}
\affiliation{Department of Theoretical Condensed Matter Physics, Universidad Autónoma de Madrid, 28049, Madrid, Spain}
\affiliation{Condensed Matter Physics Center (IFIMAC), Universidad Autónoma de Madrid, 28049, Madrid, Spain}
\affiliation{Instituto Nicolás Cabrera, Universidad Autónoma de Madrid, 28049, Madrid, Spain}
\author{Pablo Llombart}
\affiliation{Department of Theoretical Condensed Matter Physics, Universidad Autónoma de Madrid, 28049, Madrid, Spain}
\affiliation{Condensed Matter Physics Center (IFIMAC), Universidad Autónoma de Madrid, 28049, Madrid, Spain}
\affiliation{Instituto Nicolás Cabrera, Universidad Autónoma de Madrid, 28049, Madrid, Spain}
\author{Alfredo Alexander-Katz}
\affiliation{
Department of Materials Science and Engineering, Massachusetts Institute of Technology, Cambridge, MA, 02139, USA}
\author{Laura R. Arriaga}
\email{laura.rodriguezarriaga@uam.es}
\affiliation{Department of Theoretical Condensed Matter Physics, Universidad Autónoma de Madrid, 28049, Madrid, Spain}
\affiliation{Condensed Matter Physics Center (IFIMAC), Universidad Autónoma de Madrid, 28049, Madrid, Spain}
\affiliation{Instituto Nicolás Cabrera, Universidad Autónoma de Madrid, 28049, Madrid, Spain}
\author{Juan L. Aragones}
\email{juan.aragones@uam.es}
\affiliation{Department of Theoretical Condensed Matter Physics, Universidad Autónoma de Madrid, 28049, Madrid, Spain}
\affiliation{Condensed Matter Physics Center (IFIMAC), Universidad Autónoma de Madrid, 28049, Madrid, Spain}
\affiliation{Instituto Nicolás Cabrera, Universidad Autónoma de Madrid, 28049, Madrid, Spain}

\date{\today}

\begin{abstract}
Transport phenomena in complex and dynamic microscopic environments are fundamentally shaped by hydrodynamic interactions. In particular, microparticle transport in porous media is governed by the delicate interplay between particle-substrate friction and pressure forces. Here, we systematically investigate the motion of externally driven rotating magnetic microparticles near a substrate patterned with a square lattice of cylindrical obstacles, a model porous medium. Remarkably, we observe a reversal in the direction of particle translation as obstacle spacing decreases, highlighting a sensitive competition between shear-induced forward rolling and pressure-driven backward sliding due to flow-field symmetry breaking. These results demonstrate the crucial role of structured environments in determining microscale active particle transport, offering novel strategies for microfluidic design, targeted cargo delivery, and tunable active materials.
\end{abstract}

  
\maketitle

Transport is a fundamental process that spans all aspects of life, occurring across a wide range of length scales, from macroscopic systems to electrons. At the macroscopic scale, advanced sensor data and complex decision-making algorithms enable navigation through ever-changing conditions. At the quantum scale, electron transport in solids enables transformative technologies. At the microscale, controlling transport is essential for various applications, ranging from
bacterial flow management to mitigate infection risks to
the design of microfluidic systems~\cite{C8LC01323C} and microrobots for
targeted therapeutic and surgical interventions~\cite{gao2012cargo,li2017micro,palagi_bioinspired_2018,huang20233d, gao2012cargo,ahmad2021mobile,bozuyuk2024roadmap}. Yet, achieving controlled transport at the microscale remains challenging, as viscous forces dominate over inertial forces and the environment is often complex and dynamic. In these environments, characterized by a low Reynolds number, transport relies on breaking the symmetry of the surrounding flow field to generate net motion~\cite{purcell1977life,witten2020review}. An example of such symmetry breaking is an externally driven particle rotating near a substrate~\cite{sing2010controlled,steimel2014artificial}. Such particles rotating near a substrate experience roto-translational coupling due to lubrication, generating net particle translation~\cite{rogowski_symmetry_2021,morimoto2008tumbling,tierno2008magnetically,tierno2010controlled,bozuyuk2023size, chamolly_irreversible_2020, wu2022magnetic, petell2021mechanically, bozuyuk2023microrobotic, alapan_multifunctional_2020,bozuyuk2023microrobotic,qi2021quantitative,dou2021programmable}. This translational motion emerges from the competition between shear forces, resulting from the breaking of the top-down symmetry, and pressure forces caused by the breaking of the fore-aft symmetry of the rotational flow field~\cite{magrinya}. Additionally, hydrodynamic coupling of these \textit{rollers} and a single large obstacle can attract or repel them depending on the initial conditions~\cite{van_der_wee_simple_2023}, highlighting the complexity of motion in structured environments~\cite{van2025gating,jiang2022control}. This suggests that the transport of rollers in obstacle lattices is far from trivial, as hydrodynamic coupling with multiple obstacles can significantly alter the motion of a roller.

In this work, we investigate the motion of an externally driven rotating particle within a square lattice of cylindrical obstacles. The particle is actuated by a clockwise-rotating magnetic field aligned with the plane of the substrate, as shown in Fig. 1a. In the absence of the obstacle array, the roto-translational coupling induced by shear stresses propels the particle forward in the \textit{rolling} direction. In the presence of the obstacle array, we find that particle translation direction strongly depends on the ratio of the lattice spacing to particle size. When this ratio is large, particles exhibit net forward motion in the rolling direction. However, for smaller ratios, particles exhibit net backward motion in the \textit{sliding} direction. Combining experiments and simulations, we demonstrate that although the shear force varies with particle height, as observed experimentally, the direction of motion is ultimately governed by the pressure force induced by the obstacles, as shown by the simulations carried out at constant particle height. 

\begin{figure}[h!]
\centering
\includegraphics[width=0.48\textwidth]{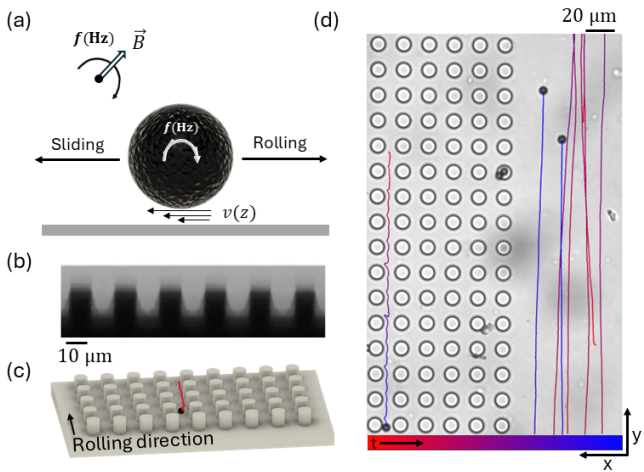}
\caption{(a) Schematic of a magnetic particle rotating near a substrate. No-slip boundary conditions cause the particle to roll in the direction of rotation. When the particle moves in  the direction opposite to its rotation, we refer to it as sliding.  (b) Confocal fluorescence microscope image showing the horizontal view of the square obstacle lattice. (c) Schematic illustration of the experimental system - a particle rotating close to a substrate patterned with a square obstacle lattice, moving along the sliding direction. (d) Representative trajectories of rotating particles inside and outside the lattice, showing motion in opposite directions. Particles are shown at their final positions, and trajectory color indicates time.}
\label{fig1}
\end{figure}

To create the model porous medium on polydimethylsiloxane (PDMS) using soft lithography, we design square lattices of cylindrical obstacles separated by a lattice spacing $l_p$, defined as the edge-to-edge distance, ranging from \SI{6} to \SI{11}{\micro\meter}~\cite{xia1998soft}. Each obstacle has a radius of \SI{5}{\micro\meter} and height of $11.5 \pm 0.5$ \SI{}{\micro\meter}, as shown in Fig.~\ref{fig1}b ~\cite{diel2020tutorial}. To enhance wetting of the PDMS substrate by the aqueous suspension, we plasma treat the PDMS for 60 s (Femto-SLS, Diener) and coat it with poly(ethylene glycol) (PEG, 6 kDa) prior to depositing the particle suspension on the substrate~\cite{long2017polyethylene}. For each experiment, we place \SI{8}{\micro\liter} of an aqueous suspension of commercial magnetic microparticles coated with biotin-PEG-COOH (5 kDa, Polysciences) on top of the PDMS structured substrate and cover it with a glass coverslip, yielding a sandwich configuration with a typical height of $\sim$ \SI{35}{\micro\meter}, which is sealed with UV-cured adhesive (NOA 68, Norland). To broaden the range of relative spacing, $l_p/R$, we use particles of radii $R$, ranging from \SI{3} to \SI{8}{\micro\meter}. Details about sample preparation can be found in the Supplemental Material (SM). Due to their density, the particles sediment to the bottom of the observation chamber between the obstacles. We then actuate the microparticles using a home-made rotating magnetic field setup, which consists of an inverted bright field optical microscope (Nikon ECLIPSE Ts2R) coupled to three orthogonal pairs of Helmholtz coils arranged along the $x$, $y$ and $z$ axis. We generate two digital sinusoidal signals phase-shifted by $\pi/2$ and send them to a digital-to-analog converter (Measurement computing USB-1208HS), which route the signals to 500 W power amplifiers (LD Systems) before reaching the coils, generating a homogeneous magnetic field of approximately 10 mT. To induce particle translation along the $y$ axis, we apply a rotating magnetic field around the $x$ axis, which is achieved by sending the sinusoidal signals to the two pairs of Helmholtz coils aligned with the $y$ and $z$ axes. 

A rotating particle suspended in an unbounded fluid generates a rotational flow without undergoing any translation. However, the proximity of the PDMS substrate along the direction parallel to the rotation axis breaks the top-down symmetry of the rotational flow, enabling particle translation. For a clockwise-rotating particle, the net shear force resulting from the gap between the particle surface and the substrate, $\delta$, drives the particle in the rolling direction, +$y$-axis. We measure the translation velocity of a rotating particle on a PDMS substrate, $v_0$, as a function of the rotational frequency of the magnetic field, $f$. As a measure of the lubrication force propelling the particle, we define the dimensionless roto-translational coupling parameter, $\xi_0 = {v_0}/{2 \pi f R}$, which quantifies the friction force propelling the particle. We obtain $\xi_0 = 0.039 \pm 0.006$  for particles with nominal radius $R$ = \SI{4.0}{\micro\meter}, $\xi_0 = 0.044 \pm 0.004$  for $R$ = \SI{2.5}{\micro\meter} and $\xi_0 = 0.048 \pm 0.005$ for $R$ = \SI{1.5}{\micro\meter} (Fig. S1, SM), indicating that in all cases the coupling is mediated by viscous forces in the lubricated friction regime~\cite{goldmans1967}. These $\xi_0$ values serve as a reference for normalizing the coupling parameter of particles within the lattice. In stark contrast to free particles, we observe that a clockwise rotating particle within an obstacle lattice translates along the sliding direction, $-y$-axis, as illustrated schematically Fig.~\ref{fig1}c and as shown in Fig.~\ref{fig1}d and Movies SM1 and SM2 .

\begin{figure}
    \centering
    \includegraphics[scale = 0.34]{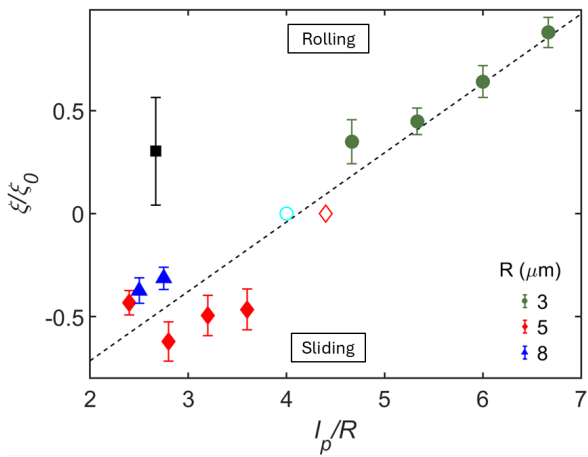}
    \caption{Normalized roto-translational coupling parameter as a function of the relative spacing in the lattice, showing the transition from sliding to rolling motion. Empty markers represent the onset of this transition. The thick discontinuous line is a linear fit to the data. The thin discontinous line represents the cutoff from rolling to sliding motion. The black square corresponds to a denser particle with nominal radius $R$ = \SI{6}{\micro\meter}.}
    \label{fig2}
\end{figure}

This reversal of the direction of motion suggests a fundamental shift in the underlying force balance within confined environments. Indeed, the sliding motion arises from net pressure forces generated by the collision of the rotational flow with the obstacles, which breaks the fore-aft symmetry of the flow, creating localized pressure gradients that propel the particle opposite to the shear-induced rolling direction. Since these pressure gradients are sensitive to confinement conditions~\cite{magrinya,gompper2010}, we track particles rotating at different frequencies within lattices of varying relative spacing, using standard tracking algorithms~\cite{crocker1996methods}. From the particle trajectories, we calculate the particle velocity, $v_p$, which is defined as positive for rolling motion and negative for sliding motion. Then, we calculate the roto-translational coupling parameter within the obstacle lattices as a function of $l_p/R$. The coupling parameter of the particle within the lattice, $\xi$, approaches that of the freely rolling particle, $\xi_0$, for the largest relative spacing. As the relative spacing decreases, the normalized coupling parameter decreases linearly, as shown in Fig.~\ref{fig2} (and Fig. S2, SM), reflecting the enhancement of pressure forces with increasing confinement. Eventually, $\xi/\xi_0$ becomes negative, indicating a transition from rolling-dominated motion to sliding-dominated motion as pressure forces overpower shear forces in more confined environments. To further test this balance between shear and pressure forces, we introduce heavier silica magnetic particles of $R = 5.4~\mu m$ and $\rho_p \sim 2.5$ g / cm$^3$ in an obstacle lattice with a small spacing, $l_p/R \sim 2.7$. In this case, the increased density results in a smaller gap between the particle surface and the substrate, leading to stronger shear forces. As predicted, we observe rolling motion at relative spacings where lighter particles exhibit sliding, as shown by the black square in Fig.~\ref{fig2} and Movie SM3. 

\begin{figure}
    \centering  \includegraphics[scale= 0.46]{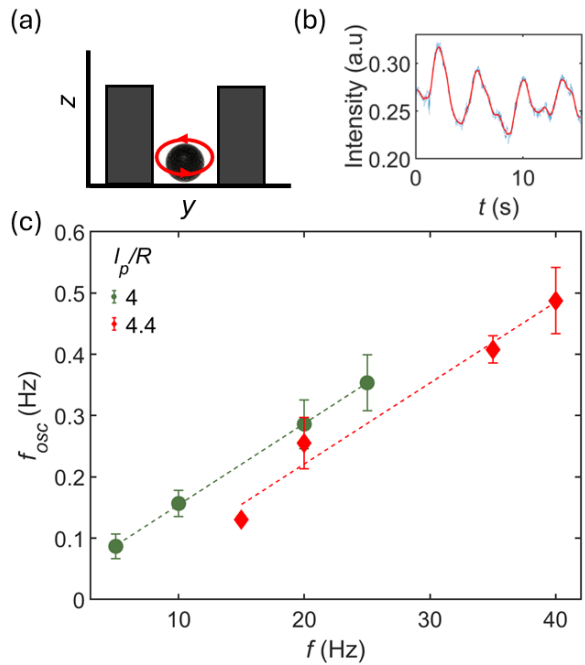}
    \caption{(a) Schematic illustration of the out-of-plane motion of a rotating particle hydrodynamically trapped. (b) Time evolution of the intensity of the brightest spot of a hydrodynamically trapped particle at $l_p/R$ = 4.0. (c) In-plane oscillation frequency as a function of the rotation frequency for $l_p/R$ = 4.0 (green circles) and $l_p/R$ = 4.4 (red diamonds). The discontinuous lines are linear fits to each data set.}
    \label{fig3}
\end{figure}

The transition from rolling to sliding motion as a function of the lattice spacing is not continuous; there exists a critical region where particles become trapped, unable to translate through the lattice. At the onset of this transition, for $l_p/R$ = 4.0 - 4.4, we observe a trapping regime where particles oscillate between rolling and sliding motion, as shown in Movie SM4, yielding $\xi/\xi_0~\sim$~0, as shown by the empty symbols in Figure~\ref{fig2}. This oscillatory behavior is consistent with an out-of-plane motion, where the particle exhibits an ellipsoidal trajectory in the vertical direction, as illustrated schematically in Figure~\ref{fig3}a. To verify this three-dimensional motion, we analyze the illumination profile of the particle by measuring the variations in the brightest pixel intensity (Figs.~S4 and S5,  SM). This allows us to extract the characteristic oscillatory pattern along the $z$-axis, shown in Figure~\ref{fig3}b, confirming that the motion includes a small but non-negligible vertical component, which changes the balance between the shear and pressure forces~\cite{demirors2021magnetic}. Furthermore, we find that the oscillation frequency in this rolling-sliding transition regime scales linearly with the particle rotation frequency, as shown in Fig.~\ref{fig3}c. This linear relationship confirms that the oscillatory motion is directly coupled to the external actuation rather than arising from spontaneous fluctuations, and suggests that the periodic vertical displacement is the driving mechanism behind the rolling-sliding transition. When the particle moves closer to the substrate, shear stresses increase, overcoming pressure forces and promoting rolling. Since shear forces scale logarithmically with the separation distance~\cite{goldmans1967}, even small changes in vertical position can significantly alter the force balance, enabling this transition to occur with vertical displacements no larger than $\sim$ \SI{1}{\micro\meter} for the highest rotation frequency experimentally accessed ($f$ = 40 Hz). 

\begin{figure}
    \centering
   \includegraphics[scale = 0.59]{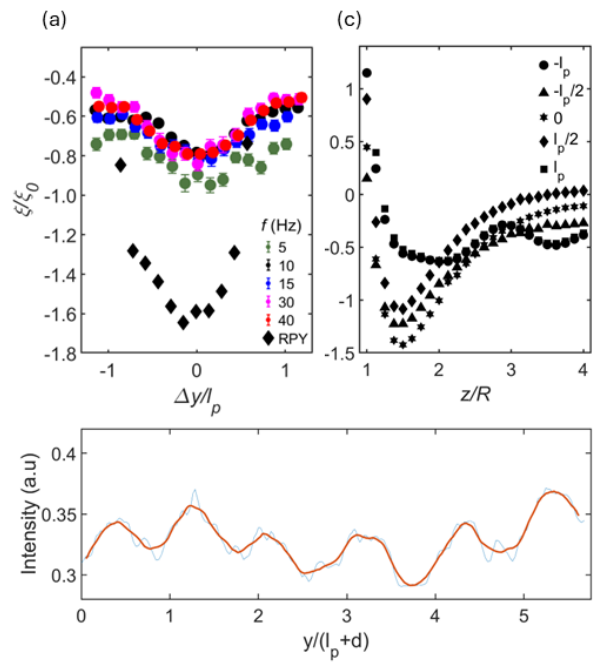}
    \caption{
    (a)  Normalized roto-translational coupling parameter as a function of the height of the rotating particle relative to the substrate from RPY simulations.  (b) Normalized roto-translational coupling parameter as a function of the distance from the center of the unit cell. The black symbols are RPY simulations performed at a fixed particle height $z/R$ = 1.75. (c)
    Trajectory of the intensity of the brightest spot of the rotating particle across the lattice along the sliding direction for $f$ = 40 Hz.}
        \label{fig4}
\end{figure}

To investigate the sliding regime that occurs at small relative spacing between obstacles, we examine the dependence of the normalized roto-translational coupling parameter on particle position within the unit cell of the lattice. We observe that regardless of the rotational frequency and the lattice spacing, particles translate faster at the center of the unit cell, as shown by the largest (negative) value of $\xi/\xi_0$ at $\Delta y/l_p$ = 0 in Fig.~\ref{fig4}a (Fig. S3, SM), respectively. This suggests that, at the center of the unit cell, shear forces decrease, pressure forces increase, or both effects act simultaneously. To assess the possible reduction of shear forces at the center of the unit cell, we measure the vertical position of the particle using the intensity profile. We find that the particle reaches the highest position at the center, as shown in Figure 4b. While this supports a reduction in shear forces at the center, it does not demonstrate that shear is the dominant factor.

To determine the role of pressure forces in promoting particle translation, we carry out Stokesian Dynamics simulations, in which the particle and the obstacle lattice are hydrodynamically coupled via the Rotne-Prager-Yamakawa (RPY) mobility tensor~\cite{rotne1969,yamakawa,Fiore_Swan_2019, ermak_mckammon}. 
In our model, the roller consists of 12 beads of radius $a$ connected by harmonic springs forming a regular dodecahedron of effective radius $R = $ 0.782$a$ ~\cite{floren_raspberry_model}, as shown in SM5. The particle is subjected to a magnetic force given by the coupling between the magnetic dipole of the particle, $\overrightarrow{m}$, and an external rotating magnetic field, $\overrightarrow{B}$, that rotates around the $x$-axis with fixed amplitude and frequency. The torque the field exerts on the particle, $\overrightarrow{\tau}=\overrightarrow{B} \times \overrightarrow{m} = B_0(0, \text{ sin} (\omega t), \text{ cos} (\omega t))\times \overrightarrow{m}$, is then distributed as forces over the beads that compose the particle. The obstacle lattice is similarly constructed of beads connected by springs, comprising a flat substrate and cylindrical obstacles of $R_o/R =$ 2 and height $h/R =$ 6.4 (Fig S6, SM). Details about the numerical simulations can be found in the SM. For a free roller, at a separation distance from the substrate of $z/R =$ 1.75, the roto-translational coupling is $\xi_0$ = 0.07~\cite{magrinya} (see Fig S7, SM). At this fixed height, the roller translates along the sliding direction when confined within the obstacle lattice, in agreement with our experimental observations (see SM6). Surprisingly, the numerical model qualitatively reproduces the non-monotonic particle velocity profile across the unit cell even at a fixed height, as shown by the black diamonds in Fig.~\ref{fig4}b. This indicates that changes in the vertical position of the particle position are not the primary contribution to the particle translational velocity. Therefore, we observe the strongest pressure force at the center of the unit cell, even at a fixed height (See fig. S8). Although both a reduction in shear and an increase in pressure are present experimentally, simulations identify pressure forces as the dominant factor dictating particle translation.   

To further investigate this behavior, we perform numerical simulations of the roller at different fixed heights, $z/R$, and various positions across the unit cell. The particle exhibits rolling motion, $\xi/\xi_0 > 0$, until it reaches $z/R \sim$ 1, beyond which it transitions into a sliding regime, $\xi/\xi_0 < 0$, independently of the particle position within the unit cell, as shown in Figure~\ref{fig4}c. In the sliding regime, as expected, the particles sliding velocity initially increases with height, reaching a maximum at $z/R \sim $ 1.75 due to the reduction of the net shear force on the roller. However, for larger particle heights, the sliding velocity decreases as the particle moves farther from the substrate. This reduction stems from the finite height of the obstacles. As the particle rises, the pressure force decreases as less of the rotational flow generated by the particle collides against obstacle surfaces. We also observe this behavior experimentally; for \SI{8}{\micro\meter} particles, the sliding velocity is lower than that of \SI{5}{\micro\meter} particles, as shown in Fig.~\ref{fig2}. This occurs because the flow field generated by the larger particles extends above the obstacles, preventing the net pressure force from scaling proportionally with particle size. These results confirm that the dependence of $\xi/\xi_0$ on particle position arises from the net pressure force induced by the fore-aft symmetry breaking of the rotational flow field within the obstacle lattice. At the cell center, the rotational flow field colliding against the obstacles is maximized, generating a reflected flow that results in a larger pressure gradient (Fig.~S8, SM). In contrast, when positioned between two obstacles, $\Delta y/l_p$ = $\pm$ 1 + 2$R_o$, the rotational flow is close to parallel to the closest two obstacles surfaces, reducing the reflected flow and, consequently, the sliding velocity. Importantly, we observe similar behavior when allowing the particle to move freely in the vertical direction (Fig.~S8, SM). Consistent with experimental observations, the particle rises as it approaches the center of the unit cell and descends as it moves away. These vertical displacements are driven by pressure forces, which possess an out-of-plane component that lifts the particle (Fig.~S9, SM). Since this lifting force depends on the rotational frequency, it gives rise to a frequency-dependent transition between rolling and sliding (SM6, SM).   

In conclusion, we demonstrated that the motion of rotating magnetic particles in patterned environments is governed by a delicate balance between shear and pressure forces. Our experiments reveal a striking transition from rolling to sliding motion as the lattice spacing decreases and pressure forces overcome shear stresses. This transition occurs through an intermediate regime in which particles exhibit oscillatory motion. This intermediate regime is enabled by three-dimensional motion, where even small vertical displacements significantly alter the force balance due to the logarithmic dependence of shear forces on separation distance. By combining experimental observations with numerical simulations, we establish that fore-aft symmetry breaking of the rotational flow field by the obstacle lattice generates position-dependent pressure forces that drive sliding motion. This effect is strongest at the unit cell center, where the collisions between fluid dragged by the rotating particle and the obstacles are maximized. These findings provide fundamental insights into the hydrodynamics of rotating particles in confined structured environments and open new avenues for designing microfluidic systems with controlled particle transport. The ability to reverse particle motion through simple geometric modifications offers promising applications in microfluidic sorting and controlled self-steering of microscopic cargo carriers.

\begin{acknowledgments}
\noindent The authors acknowledge financial support by MCIN/AEI/10.13039/501100011033/ for all grants listed next: JLA and LRA for PID2022-143010NB-I00 and CEX2023-001316-M, also supported by "ERDF A way of making Europe"; JLA for RYC2019-028189-I and CNS2023-145447, also supported by "European Union NextGenerationEU/PRTR"; LRA for CNS2023-145460, also supported by "European Union NextGenerationEU/PRTR"; GG for PRE-2021-099492, also supported by "ESF Investing in your future". AAK is grateful to MISTI Spain for funding and to the Michael and Sonja Koerner chair for financial support as well. The authors thank Prof. A. Hidalgo for providing access to the plasma etcher and Profs. J.V. Alvarez and E. Lauga for insightful discussions. 
\end{acknowledgments}

\bibliographystyle{apsrev4-1}
%

\end{document}